\documentclass{aa}  
\usepackage{graphicx}
\usepackage{txfonts}

\def\gap{\;\rlap{\lower 2.5pt
 \hbox{$\sim$}}\raise 1.5pt\hbox{$>$}\;}
\def\lap{\;\rlap{\lower 2.5pt
   \hbox{$\sim$}}\raise 1.5pt\hbox{$<$}\;}
\def\gsim{\;\rlap{\lower 2.5pt
 \hbox{$\sim$}}\raise 1.5pt\hbox{$>$}\;}
\def\lsim{\;\rlap{\lower 2.5pt
 \hbox{$\sim$}}\raise 1.5pt\hbox{$<$}\;}
\def\msun{{\rm\,M_\odot}}

\def\cm{{\rm\,cm}}
\def\sec{{\rm\,s}}

\def\sr{{\rm\,sr}}

\def\MeV{{\rm\,MeV}}

\def\cm{{\rm\,cm}}

\def\kpc{{\rm\,kpc}}

\def\GeV{{\rm\,GeV}}

\def\sec{{\rm\,s}}
\def\sr{{\rm\,sr}}

\def\spose#1{\hbox to 0pt{#1\hss}}
\def\lta{\mathrel{\spose{\lower 3pt\hbox{$\mathchar''218$}}
     \raise 2.0pt\hbox{$\mathchar''13C$}}}
\def\gta{\mathrel{\spose{\lower 3pt\hbox{$\mathchar''218$}}
     \raise 2.0pt\hbox{$\mathchar''13E$}}}
\newcommand{\beq}{\begin{equation}}
\newcommand{\eeq}{\end{equation}}
\newcommand{\be}{\begin{equation}}
\newcommand{\ee}{\end{equation}}

\newcommand{\ls}{\mathrel{\raise1.16pt\hbox{$<$}\kern-7.0pt 
\lower3.06pt\hbox{{$\scriptstyle \sim$}}}}         
\newcommand{\gs}{\mathrel{\raise1.16pt\hbox{$>$}\kern-7.0pt 
\lower3.06pt\hbox{{$\scriptstyle \sim$}}}}         

\long\def\comment#1{}

\def\fun#1#2{\lower3.6pt\vbox{\baselineskip0pt\lineskip.9pt
  \ialign{$\mathsurround=0pt#1\hfil##\hfil$\crcr#2\crcr\sim\crcr}}}
\def\lap{\mathrel{\mathpalette\fun <}}
\def\gap{\mathrel{\mathpalette\fun >}}
\newcommand{\ba}{\begin{eqnarray}}
\newcommand{\ea}{\end{eqnarray}}

\def\rsun{{R_\odot}}    

\def\kms{$\rm km\;s^{-1}$}

\def\cms{$\rm cm^3\;s^{-1}$}

\def\phcms{$\rm ph\;cm^{-2}\;s^{-1}$}

\begin{document}
   \title{Could the Fermi-LAT detect $\gamma$-rays from dark matter
   annihilation in the dwarf galaxies of the Local Group?}


   \author{L. Pieri\inst{1}\inst{2}\inst{3}
          \and
          A. Pizzella\inst{3}
          \and
          E. M. Corsini\inst{3}
          \and
          E. Dalla Bont\'a\inst{3}
          \and
          F. Bertola\inst{3}
          }

   \offprints{L. Pieri}

   \institute{Consorzio Interuniversitario di Fisica Spaziale, 
Villa Gualino, Viale Settimio Severo, 63, I-10133 Torino, Italy \\
              \email{lidia.pieri@oapd.inaf.it}
        \and
            Istituto Nazionale di Fisica Nucleare - Sezione di Padova,  
  via Marzolo 8, I-35131 Padova, Italy 
         \and
             Dipartimento di Astronomia, Universit\`a di Padova,  
  vicolo dell'Osservatorio~3, I-35122 Padova, Italy   
             }


 
  \abstract
   {The detection of $\gamma$-rays from dark matter (DM) annihilation is
among the scientific goals of the Fermi Large Area Telescope
(formerly known as GLAST) and Cherenkov telescopes.}
   {In this paper we investigate the existence of realistic chances of such
a discovery selecting some nearby dwarf spheroidal galaxies (dSph) as a target,
and adopting the DM density profiles derived from both astronomical 
observations and $N$-body simulations.
We also make use of recent highlights about the presence of 
black holes and of a population 
of sub-subhalos inside the Local Group (LG) dwarfs for
boost factor studies.}
   {We study the detectability with the Fermi-LAT of the $\gamma$-ray flux from DM
annihilation in four of the nearest and highly DM-dominated dSph
galaxies of the LG, namely Draco, Ursa Minor, Carina, and
Sextans, for which the state-of-art DM density profiles were available.  
We assume the DM is made of Weakly Interacting Massive Particles
such as the Lightest Supersymmetric Particle (LSP) and compute the
expected $\gamma$-ray flux for optimistic choices of the unknown 
underlying particle physics parameters.
We then compute the boost factors due to the presence of DM clumps and of a
central supermassive black hole.
Finally, we compare our predictions with the Fermi-LAT sensitivity maps.}
   {We find that the dSph galaxies shine above the Galactic 
smooth halo: e.g., the Galactic halo is brighter than the Draco dSph 
only for angles smaller than 2.3 degrees above the Galactic Center. 
We also find that the presence of a cusp or a constant density core in the DM
mass density profile does not produce any relevant effect in the
$\gamma$-ray flux due to the fortunate combination of the geometrical 
acceptance of the Fermi-LAT detector
and the distance of the galaxies and that no significant enhancement is given by the presence of a central black hole or 
a population of sub-subhalos.}
 { We conclude that, even for the most optimistic scenario of particle physics,
the $\gamma$-ray flux from DM annihilation in the dSph galaxies of the LG
would be too low to be detected with the Fermi-LAT.}


   \maketitle
%

\section{Introduction}
\label{sec:introduction}

Since the first evidence of the presence of dark matter (DM) in the
universe, scientists have worked to understand its nature and
distribution. This investigation involves different fields of research
such as particle physics, cosmology and, observational astronomy
(e.g. \cite{Bahcall1999, Spergel2003}).

The Fermi Large Area Telecope (Fermi-LAT) 
will test theories in which DM candidates are the Lightest
Supersymmetric Particles (LSPs) such as the neutralinos, arising in
Supersymmetric extensions of the Standard Model of particle physics
(SUSY), or the Lightest Kaluza-Klein Particles (LKKPs) such as the
$B^{(1)}$s, first excitation of the hypercharge gauge boson in
theories with Universal Extra Dimensions
(see \cite{Bergstrom2000, Bertone2005}, and references therein).
Typical values for the mass of these candidates range from about 50
GeV up to several TeV.

Cosmological models, mainly based on $N-$body simulations in a
$\Lambda$-Cold Dark Matter (CDM) framework, successfully reproduce
relevant characteristics of the universe such as the cosmic microwave
background anisotropy and the large scale structure of the
universe. They also predict well-defined properties of DM haloes,
whose radial mass density distribution follows a
universal law, and it is described by a steep power law for a wide range
of masses ranging from dwarf galaxies to galaxy clusters
(see, e.g., \cite{Navarro1996, Navarro1997, Navarro2004, Moore1998, Moore1999,
Diemand2005}).
However, the astronomical community is still debating whether DM
haloes are characterized by a central density cusp. In fact, haloes
with a constant density core are in most cases preferred to account
for the observed kinematics of galaxies (see \cite{Binney2004} for a
review).

Generally speaking, the uncertainty in the choice of the density
profile can result in several order of magnitude of indetermination
on the $\gamma$-ray flux prediction, which already suffers from the
high uncertainties arising by the unknown underlying particle physics
(\cite{Fornengo2004}).
For this reason it would be important to derive the DM
density profile of galaxies directly from the available kinematic
data. Although data-sets for the very inner part of the galaxies are
scarce and affected by large errors, the situation is not better in
$N-$body simulations, whose resolution goes down to 0.05 times
the virial radius at most.
Using real data we have the advantage of deriving a flux
prediction which takes into account the peculiarity of each galaxy,
without any model-dependent generalization which would arise the
astrophysical uncertainties.

The expected $\gamma$-ray flux at the telescope from a given source is
directly proportional to the DM density squared along the line-of-sight (LOS), and inversely proportional to the square of its distance.
The best targets are therefore nearby dense objects such as the local
dwarf spheroidal galaxies (see \cite{Mateo1998} for a review).
Indeed, in the last decade, the large collecting area of the 8-m class
telescope and the use of multi-fiber spectrographs allowed astronomers
to obtain high-resolution spectra of a large number of stars. This
made possible to isolate the galaxy member stars, to measure their
radial velocity with an accuracy of few \kms\ and to build accurate
dynamical models of a number of such systems
(see \cite{Gilmore2007} and references therein).

Annihilation of $\gamma$-rays in dSph galaxies would give a clean
signal because of the absence of high astrophysical uncertainties in
modeling the expected background and could hopefully be detected with
upcoming experiments like the Fermi-LAT. Many authors studied the feasibility
of such a detection, using a large variety of cuspy and cored universal
density profiles, reflecting the theoretical as well as the
experimental uncertainties.
Different works (\cite{Baltz2000,Tyler2002,Peirani2004,Pieri2004,Bergstrom2006})
found that only the presence 
of a spike and/or an enhancement due to clumpiness and/or a more
favourable combination of the unknown particle physics parameters
could make the Draco dwarf galaxy observable with the Fermi-LAT.
\cite{Strigari2007} are optimistic about the 
detection of Draco with the Fermi-LAT
in 5 years. They adopted a King profile (\cite{King1966}) to derive the
surface density of the stellar luminosity. This was deprojected and
converted into the stellar mass density by adopting the typical range
of the mass-to-light ratio of dSphs. The luminous mass they derived is at the
very least one order of magnitude below the mass of the DM halo.
For the halo they assumed a NFW mass density profile (\cite{Navarro1996,
Navarro1997}) whose free degenerate parameters were constrained by
marginalizing over the stellar velocity dispersion anisotropy
parameter.
\cite{Colafrancesco2007} showed how diffuse radio emission would
actually be a more promising process to look at in order to detect a DM
signal. They also claimed that he presence of a supermassive black
hole (SBH) at the centre of Draco, which could enhance the
$\gamma$-ray signal up to detection, is not actually excluded by
experiments.
Detection of annihilation $\gamma$-rays from Draco has been recently
excluded by \cite{Sanchez2007} through the use of density profiles
which are compatible with the latest observations. 

In this paper we use the latest available astrophysical measurements
for four of the nearest and highly DM-dominated dSph galaxies of the
Local Group, namely  Draco, Ursa Minor, Carina, and Sextans. 
to compute the expected $\gamma$-ray flux from DM annihilation. \\

In Sect.~\ref{sec:flux} the most optimistic particle physics scenarios
and the DM density profiles derived both from the available kinematic
measurements and from $N$-body simulations 
are used to predict the expected $\gamma$-ray flux from
DM annihilation in Draco, Ursa Minor, Carina, and Sextans. In
Sect.~\ref{sec:glast} the predicted flux is compared with the experimental
sensitivity of the Fermi-LAT. The presence of DM clumps and a central SBH
could enhance the $\gamma$-ray flux. But their effects have to be
rescaled for the limits imposed on the extragalactic $\gamma$-ray
background (EGB) by the Energetic Gamma-Ray Experiment (EGRET) and on the
$\gamma$-ray flux in Draco by the measurements of the Major Atmospheric
Gamma-Ray Imaging Cherenkov (MAGIC) telescope. Our conclusions are
given in Sect.~\ref{sec:conclusions}. \\

The main differences with the other papers that already discussed 
this argument are the following: we show that, even adopting 
a very lucky case for the unknown particle physics sector, the expected
flux from the DM halo is about two orders of magnitude below the detectability
limit of the Fermi-LAT experiment; we also show how the use of a cored or a
cuspy profiles does not produce any relevant effect in the expected
$\gamma$-ray flux, because of a combination of the galaxy distance  
and the angular acceptance of the Fermi-LAT;
we then show that the current limits on the mass of the supermassive black hole
(SMBH)
inside Draco lead to an insignificant boost factor due to the presence
of such a SMBH; and we numerically compute the boost factor 
due to the presence of a population of subhaloes inside the dwarf galaxies,
limiting the possible range of models for the sub-subhalo structure making use
of the constraints imposed by the EGRET extragalactic measurements;
the boost factor due to the presence of sub-subhaloes is computed 
in two ways: first, we obtain it integrating over the whole volume of 
the galaxy, as done, e.g., in \cite{Strigari2007}. 
This gives the correct boost factor when considering cosmological haloes.  
But, if we consider the closer dwarf galaxies, we have to take into
account that only the very inner part of the galaxy is observed within
the angular resolution of the instrument.
We therefore also compute the angular dependence of the boost factor due to
sub-subhaloes. Although we find a huge enhancement of the
flux far from the galaxy center, there is actually no enhancement
along the LOS pointing toward the galaxy center.
As a last improvement with respect to the other papers, 
we compare our predictions with an recently released 
detectability map for the Fermi-LAT
which takes into account the response of the detector to different
energies and incidence angles, as far as effective energy and
angular resolution are concerned.

\section{$\gamma$-ray flux from dark matter annihilation}
\label{sec:flux}

The $\gamma$-ray flux $\Phi_\gamma$ from DM annihilation can be
factorized into a term $\Phi_{\rm PP}$ involving the particle
physics and a term $\Phi_{\rm cosmo}$ where astrophysics, cosmology,
and experimental geometry play the main role. It is
\begin{equation}
\Phi_\gamma (E_\gamma, \psi, \Delta \Omega) = 
\Phi_{\rm PP} (E_\gamma) \times \Phi_{\rm cosmo} (\psi, \Delta \Omega).
\label{eq:phi}
\end{equation} 

The particle physics factor is given by
\begin{equation}
\Phi_{\rm PP}(E_\gamma) =  
  \frac{1}{4 \pi} \frac{\sigma_{\rm ann} v}{2 m^2_{\rm DM}}  
  \sum_{\rm f} B_{\rm f} \int_{\rm E_{th}} 
  \frac{d N^{\rm f}_\gamma}{d E_\gamma} dE \, ,
\label{eq:phisusy}
\end{equation}
where $m_{\rm DM}$ is the DM particle mass, $\sigma_{\rm ann}$ is the
annihilation cross section, and $v$ is the relative
velocity. $\sigma_{\rm ann}v$ determines the number of
annihilations. $B_f$ is the branching ratio into a final state $f$. It
represents the probability that the final state $f$ is the result of
one annihilation. ${d N^{\rm f}_\gamma}/{d E_\gamma}$ is the yield of
photons produced by the final state $f$ in one annihilation, and
$E_{\rm th}$ is the threshold energy above which the flux is computed.
So far no assumptions have been made on the nature of the DM particles. For
a complete set of the allowed values for the previous parameters the
reader is referred to \cite{Fornengo2004}.

The astrophysical/cosmological factor is given by
\begin{equation}
\Phi_{\rm cosmo}(\psi, \Delta \Omega) = 
  \int \int_{\Delta \Omega} d \phi  d \theta  
  \int_{\rm LOS} d\lambda 
  \left [ \frac{\rho_{\rm DM}^2 (r)} {\lambda^{2}} J \right] \, ,
\label{eq:phicosmo}
\end{equation}
where $\psi$ is the angle of view from the halo centre which defines
the LOS and $\Delta \Omega$ corresponds to the angular
resolution of the instrument. It is in principle function of the
photon energy, $E_\gamma$, though in the following we will assume for
simplicity $\Delta \Omega= 10^{-5}$ sr (corresponding to a cone of
view with angular opening of 0.1 degree along the LOS). This
corresponds to the geometrical acceptance of the Fermi-LAT
detector. Indeed, we made the simplifying assumption that the angular
resolution of the Fermi-LAT is $0.1$ degree over the entire energy range. This
resolution is reached only for 10 GeV photons and for incidence angles
less than $50^\circ$. $\lambda$ is the coordinate along the LOS, and
$r$ the radial coordinate inside the halo. $\rho_{\rm DM}(r)$ is the
DM mass density profile, which is a factor of primary importance when
deriving the $\gamma$-ray flux from DM annihilation in a given
source. Finally, $J(x,y,z|\lambda,\theta, \phi)$ is the Jacobian
determinant.

\subsection{The particle physics factor}
\label{sec:pp}

\begin{figure}
   \includegraphics[width=0.5\textwidth]{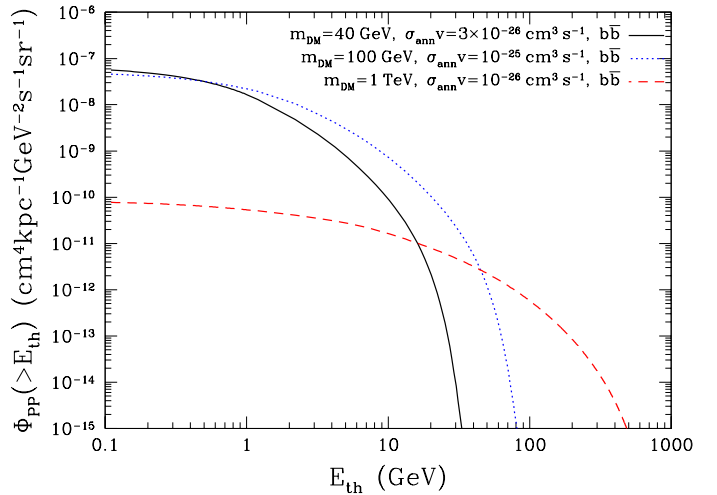}
\caption{Behaviour of $\Phi_{\rm PP} (> E_{\rm th})$
as a function of $E_{\rm th}$ for different models of the DM particle,
computed for a:
40 GeV (solid line), 100 GeV (dotted)
and 1 TeV (dashed) DM particle 
annihilating into $b \bar b$. 
The value of $\sigma_{\rm ann} v$ has been chosen as representative
of the best value for that mass,
as it is computed with DARKSUSY and allowed by WMAP+SDSS
measurements (see details in the text).}
\label{fig:phisusy}
\end{figure}

In Fig.~\ref{fig:phisusy} we draw the factor $\Phi_{\rm PP}$ integrated
above a threshold energy $E_{\rm th}$ as
a function of $E_{\rm th}$. Shown in the plot is 
the result of the computation for a 40 GeV, 100 GeV 
and 1 TeV DM particle annihilating into quarks $b \bar b$. 
For the photon yields we have used the parametric formula from 
\cite{Fornengo2004} and introduced the pion bump feature at low energies. 
For each mass, we adopted the most optimistic value for $\sigma_{\rm ann} v$, 
as it is computed with DARKSUSY (\cite{Gondolo2004}) and allowed by WMAP+SDSS
measurements (e.g, see Fig.~3 in \cite{Pieri2007}). However,
really few models lie in those very fortunate parts of the phase-space.
In details, we used  $\sigma_{\rm ann} v = 3 \times 10^{-26}$ \cms for a 
40 GeV DM particle,  $\sigma_{\rm ann} v = 10^{-25}$ \cms for the 100 GeV 
particle case and  $\sigma_{\rm ann} v = 10^{-26}$ \cms for the 1 TeV one.

We do not consider here the result of 
\cite{Bringmann2007}, who pointed out how so far ignored effects 
of electromagnetic
radiative corrections to all leading annihilation processes in the
Minimal Supersymmetric Model or in the Minimal SUperGRAvity mediated
supersymmetry breaking scenarios can induce a $\gamma$-ray flux
enhancement up to three-four order of magnitudes with respect to the
$\gamma$-ray secondary flux produced in the annihilation cascade, when
integrating over energies greater than 60 \% of $m_{\rm DM}$, even
for LSP masses well below the TeV scale. 
A careful study of the effect of internal bremsstrahlung would be interesting
for instruments with higher sensitivity at higher energies, such as 
Cherenkov telescopes, and is beyond the goal of this paper. \\

In the following, we will refer to a 40 GeV DM particle with 
$\sigma_{\rm ann} v$ = $3 \times  10^{-26}$ \cms, annihilating into $b \bar b$
as to our best case scenario
when studying the maximal $\gamma$-ray flux prediction integrated 
above 100 MeV. 
The reader must be aware that this is not the most likely model, and
that the real flux could be orders of magnitude smaller.

\subsection{The astrophysical/cosmological factor}
\label{sec:cosmo}

In this section we derive the value of $\Phi_{\rm cosmo}$ for four
dSph galaxies of the Local Group both from the state-of-art DM density
profiles available in literature and from CDM $N$-body simulations.
Their positions, masses, and distances are reported in Table~\ref{tab:sample}.

\begin{table}
\centering
\caption{The sample galaxies. Distances and masses are taken 
from \cite{Gilmore2007}} 
\begin{tabular}{lrrrr} \hline 
Object & $l$ & $b$ & Mass & Distance\\
     &deg& deg& M$_\odot$ & kpc  \\ \hline 
Draco     &  86.37 &  34.77 & $2.8 \times 10^7$& 80 \\ 
Ursa Minor& 104.98 &  44.86 & $1.3 \times 10^7$& 66 \\ 
Carina    & 260.09 & -22.28 & $1.3 \times 10^7$& 87 \\ 
Sextants  & 243.42 &  42.16 & $1.9 \times 10^7$& 80 \\ 
\hline 
\end{tabular} 
\label{tab:sample} 
\end{table}

\begin{itemize}

\item 
{\bf Draco:} \cite{Gilmore2007} calculated the DM density radial
profile of the Draco dwarf galaxy. It was derived by
\cite{Wilkinson2004} from the radial profiles of the velocity
dispersion and surface brightness by solving the Jeans equations under
the assumption of isotropic orbital structure.
The velocity dispersion radial profile extends out to about 35 arcmin
from the centre (corresponding to 0.8 kpc). It is characterized by
an almost constant value of about 13 \kms , with a decrease to about 5
\kms\ at the last observed radius. The available data allowed to
derive the mass density profile of the DM between about 0.1 and 0.5
kpc from the galaxy centre. The mass density increases out to the
innermost observed point (Fig.~\ref{fig:densitydraco}).

\begin{figure} 
   \includegraphics[width=0.5\textwidth]{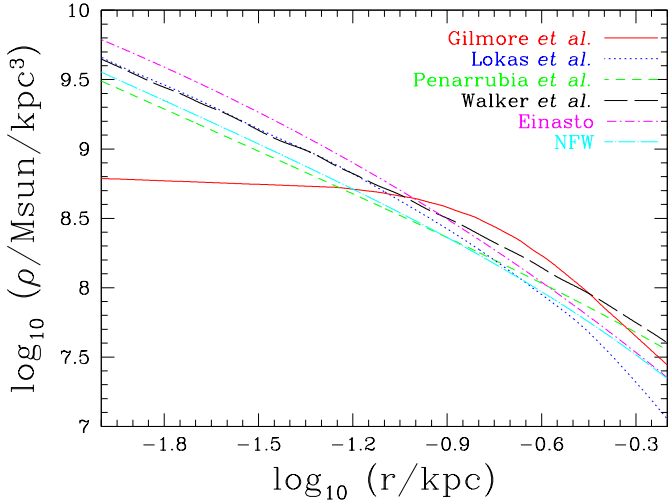}
\caption{The radial profile of the DM mass density in  
  Draco as derived by \cite{Gilmore2007} (solid line),
  \cite{Lokas2005} (dotted), \cite{Walker2007} (long-dashed), and 
\cite{Penarrubia2008} (dashed). Also shown are the 
density profiles derived 
from numerical simulations, namely the standard NFW (long-long-dashed) and the 
Einasto (long-dot-dashed) radial profiles. }
\label{fig:densitydraco}
\end{figure}

Recently, an independent mass density profile for Draco has been
obtained by \cite{Penarrubia2008}. They used the data by
\cite{Wilkinson2004} and \cite{Munoz2006} to reconstruct the mass 
distribution of the galaxy. They assumed that the galaxy is composed
by a luminous component described by means of a King model
\cite{King1966} and a DM component described by a NFW model. In this way
they derived the concentration parameter of the DM halo component
directly from a fit to the data, instead of assuming it from the CDM
cosmology. A total mass of $6.2 \times 10^9$ M$_\odot$ was found,
which is somehow larger than expected for dSph galaxies
(\cite{Mateo1998}). This DM density profile is shown in
Fig.~\ref{fig:densitydraco}. 
The same procedure was applied to the other galaxies we analyze. Yet,
as we will show in the following, this profile implies a more
pessimistic $\gamma$-ray flux prediction. Since we are interested in
the most optimistic scenarios which possibly could lead to detection,
we will consider the DM profiles derived by \cite{Penarrubia2008} only
in the case of Draco to derive the model uncertainty, while we will no
further consider it for the other galaxies.

A third DM profile for Draco was obtained by
\cite{Lokas2005}. They used the data-set by \cite{Wilkinson2004} and
assumed a S\'ersic law (\cite{Sersic1968}) to describe the distribution
of the luminous component. Concerning the DM density distribution,
they assumed a modified NFW with an inner cusp and an exponential
cut-off to take into account a possible tidal stripping in the outer
regions of the galaxy. A tidal interaction does not affect the DM mass
density profile in the centre, but produces a mass loss for radii
larger than the so-called break radius (\cite{Kazantzidis2004}). 
\cite{Lokas2005} break the degeneracy between the mass distribution
and velocity anisotropy by fitting both the LOS velocity dispersion
and kurtosis profiles. They found a total mass of $7 \times 10^7$
M$_\odot$. The corresponding radial profile of the DM mass density
profile is shown in Fig.~\ref{fig:densitydraco}.

Finally, we show in Fig.~\ref{fig:densitydraco} the density
profile obtained by \cite{Walker2007} adopting a one-component King
profile and a NFW profile with constant anisotropy parameter
for the luminous and DM components, respectively. 

A cored profile seems to be preferred for
the DM mass density when no parametric function is imposed in fitting
the data. A primordial density core would exclude a pure CDM 
scenario, rather pointing toward a warm dark matter particle. Yet, there
are different studies about the possibility of dynamically remove the 
CDM cusp in the dwarf galaxies, 
involving phenomena such as stellar feedback (e.g., \cite{Mashchenko,Read}) or
dynamical friction of DM/baryons subhaloes (\cite{Romano}).
The reader should be aware that the topic is still controverse 
(see, e.g, \cite{Gnedin}) and there is no 
univoque consensus about the realistic possibility that CDM cusps in 
dwarfs may be reduced to a core. \\
In lack of negative evidences we keep on using cored profiles associated
with CDM particles in our discussion. \\
Although the main aim of the present paper is to present results based on 
density profiles directly inferred by astronomical data, it is worth 
superimposing on Fig.~\ref{fig:densitydraco} the density profiles derived from 
numerical simulations. 
\cite{Stadel2008} have recently obtained from $N$-body 
simulations a best fit to a MW-sized halo which is a simple power law 
in $dlog(\rho)/dlog(r)$ (called the Stadel \& Moore profile -S\&M-). 
In the lack of halo mass scaling relations for the parameters of the 
S\&M profile, we show only the NFW and the Einasto profile computed 
for a $10^9 \msun$ halo at a distance of 80 kpc, with the concentration 
parameter given by \cite{Kuhlen2008} ($c=20.2$, $r_s=1.02$ kpc, 
$\rho_s^{NFW}=3.56 \times 10^7 \msun \kpc^{-3}$, 
$\rho_s^{Einasto}=8.9 \times 10^6 \msun \kpc^{-3}$.)

\begin{figure}
   \includegraphics[width=0.5\textwidth]{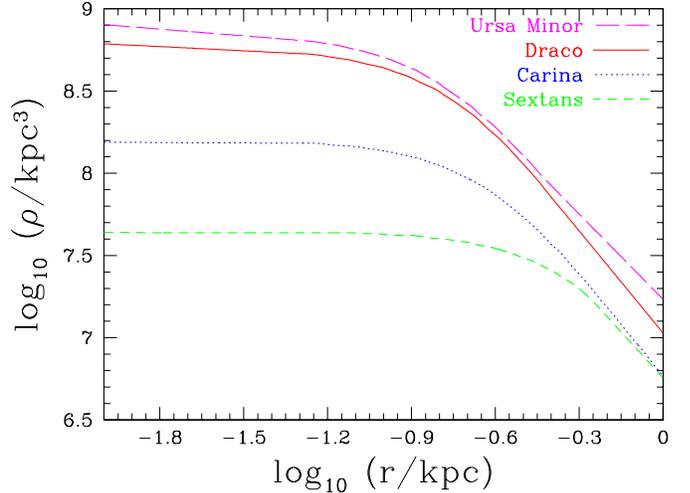}
\caption{The radial profile of the DM mass density in  
  Ursa Minor (long-dashed line), Draco (solid line), Carina (dotted
  line), and Sextans (short-dashed line) as derived by
  \cite{Gilmore2007}.}
\label{fig:density}
\end{figure}

\item
{\bf Ursa Minor:} The DM mass density profile for the Ursa Minor dwarf
galaxy was taken from \cite{Gilmore2007}. It was derived by
\cite{Wilkinson2004} from the radial profiles of the velocity
dispersion and surface brightness by solving the Jeans equations under
the assumption of isotropic orbital structure.  
The velocity dispersion radial profile extends out to 45 arcmin from
the centre (corresponding to 0.9 kpc). It is characterized by constant
value of about 12 \kms , showing a sharp drop to about 2
\kms\ only at the farthest observed radius. The data allowed to derive 
the DM density distribution in the radial range between about 0.1 and
0.5 kpc. It is similar to that of Draco (Fig.~\ref{fig:density}).

Ursa Minor was also studied in \cite{Strigari2007}, using the data
by \cite{Palma2003}. The light distribution was derived considering
a two-component, spherically-symmetric King profile. They
used a Jeans equations and adopted a NFW DM halo to the derive the
radial profile of the velocity dispersion fitting the data.
 The anisotropy
parameter $\beta$ has been empirically set to the value of $0.6$.  
In this paper we consider the two NFW models of the DM density profile
given by \cite{Strigari2007}: Model A has $r_s = 0.63 \kpc$ and 
$\rho_s = 10^8 \msun \kpc^{-3}$,
Model B has $r_s = 3.1 \kpc$ and $\rho_s = 10^7 \msun
\kpc^{-3}$. \\

\item
{\bf Carina:} \cite{Gilmore2007} calculated the DM density radial
profile of the Carina dwarf galaxy. It was derived from the radial
profiles of the velocity dispersion and surface brightness by solving
the Jeans equations under the assumption of isotropic orbital
structure.  
The available measurements extend out to about the tidal radius of the
galaxy, which corresponds to about 25 arcmin (corresponding to 0.6
kpc). The velocity dispersion is characterized by a constant value of
about 8 \kms . The DM mass density profile is derived out to 60 pc
from the centre and it shows a constant density core
(Fig.~\ref{fig:density}). \\

\item
{\bf Sextans:} The DM mass density profile for the Sextans dwarf
galaxy was taken from \cite{Gilmore2007}. It was derived by
\cite{Wilkinson2006} from the radial profiles of the velocity
dispersion and surface brightness measured by
\cite{Kleyna2004} by solving the Jeans equations under the assumption
of isotropic orbital structure. 
The velocity dispersion radial profile extends out to about 47 arcmin
from the centre (corresponding to 1.1 kpc). It is characterized by
a constant value of about 8 \kms , with a possible decrease to about 3
\kms\ at the last observed radius. The available data allowed to
derive the mass density profile of the DM between about 0.2 and 0.8
kpc from the galaxy centre. It shows a constant density core
(Fig.~\ref{fig:density}).

\end{itemize}

\begin{figure}
   \includegraphics[width=0.5\textwidth]{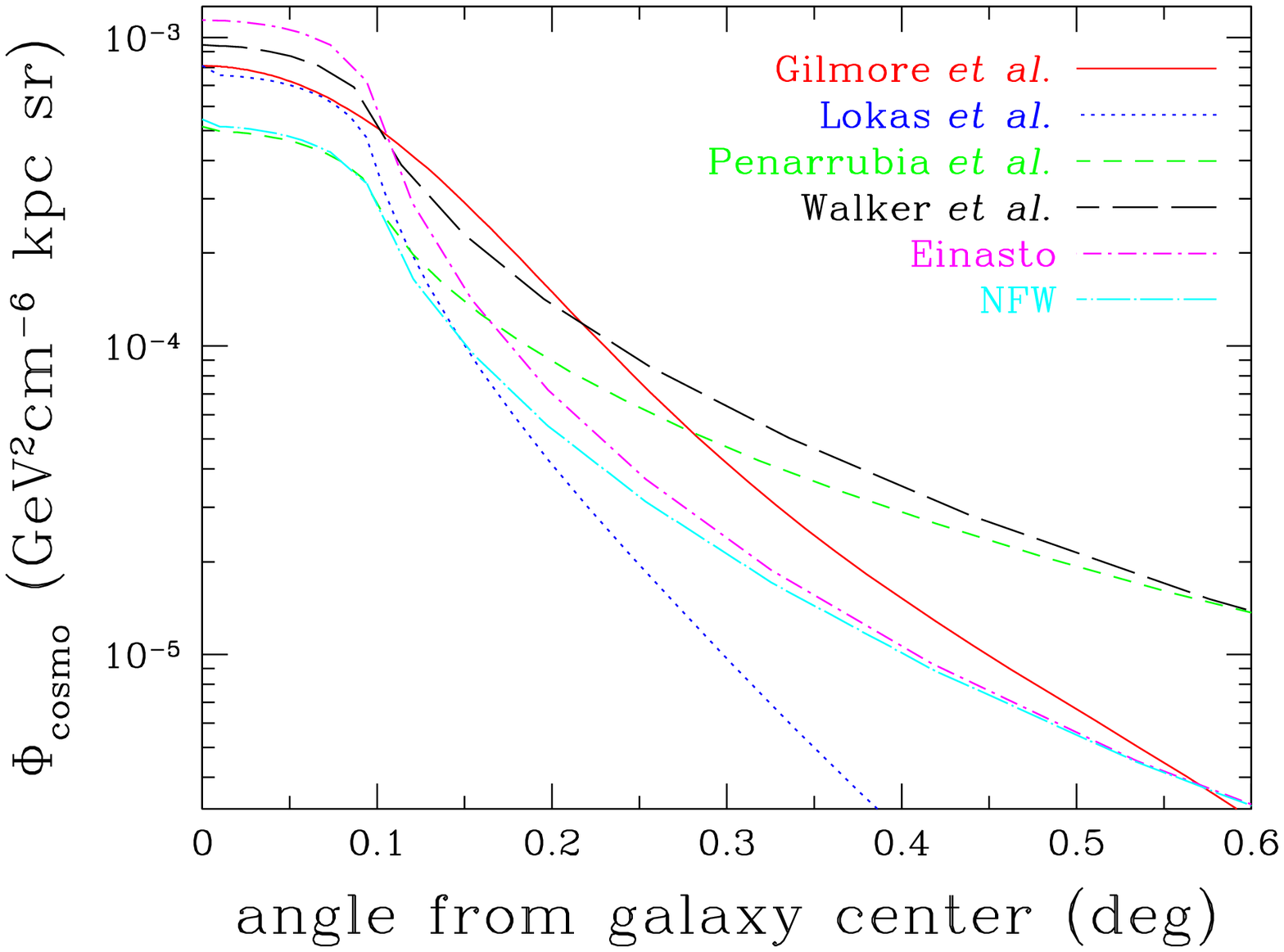}
\caption{The astrophysical/cosmological contribution $\Phi_{\rm
  cosmo}$ to the $\gamma$-ray flux derived as a function of the
  angular distance from the galaxy centre for Draco from the DM mass
  density radial profiles by \cite{Gilmore2007} (solid line),
  \cite{Lokas2005} (dotted),\cite{Walker2007} (long-dashed), and 
\cite{Penarrubia2008} (dashed).  Also shown are the results obtained from
the density profiles derived 
from numerical simulations, namely the NFW standard (long-long-dashed) and the 
Einasto (long-dot-dashed) profiles.}
\label{fig:phicosmo_draco}
\end{figure}

We would like to underline that the experimental results which we
will use in our analysis give both cuspy and cored profiles. Although CDM 
simulations predict only cuspy haloes, we will keep on considering 
cored profiles and CDM particle because there may be
mechanisms of gravitational heating of the dark matter by baryonic
components which could reconcile observations
of cored density profiles with the central density cusps of the CDM
predictions. \\
Eq.~\ref{eq:phicosmo} has then been integrated along the LOS adopting the DM
density profiles we derived for each dSph galaxy. The result of this
integration for the four different profiles inferred by the data as well as for the two profiles derived from numerical simulations for the Draco 
dSph galaxy
is found in Fig.~\ref{fig:phicosmo_draco}. The behaviour of these
curves reflects the different DM distributions shown in
Fig.~\ref{fig:densitydraco}. The fit to the data was obtained in the
radial interval between 80 and 630 pc. The DM mass density profiles
are extrapolated in the innermost and outermost galaxy regions. At
large radii the DM mass density derived by \cite{Penarrubia2008}, who
adopted a NFW density profile with no tidal disruption, is higher than
that by \cite{Lokas2005} and \cite{Gilmore2007}. Indeed,
it is comparable to the result obtained by \cite{Walker2007}, 
who also fit a NFW profile.
Actually, the total mass derived by \cite{Penarrubia2008} and 
\cite{Walker2007} is larger than the one typically found for this kind
of galaxies (\cite{Mateo1998}). This behaviour reflects in the
radial trend of the corresponding $\Phi_{\rm cosmo}$
(Fig.~\ref{fig:phicosmo_draco}), which is
higher at large radii with respect to those based on the results by
\cite{Lokas2005} and \cite{Gilmore2007}.
The results based on the \cite{Walker2007} profile give a higher value of 
 $\Phi_{\rm cosmo}$ in the inner galaxy with respect to the  
\cite{Penarrubia2008},
since the former predict a larger mass content at small radii 
(see Fig.~\ref{fig:densitydraco}).

The low DM mass density observed at large radii in the NFW profile of
\cite{Lokas2005} is due to the mass stripping induced by a tidal
interaction. Their DM density is higher in the centre, while the cored
density profile by \cite{Gilmore2007} allocates more mass at
intermediate radii (Fig.~\ref{fig:densitydraco}). Though biased by the
different derived masses, this effect is due to mass conservation
since the two models have about the same tidal radius. The DM mass
density profile by \cite{Gilmore2007} gives a larger $\Phi_{\rm
cosmo}$ for radii larger than 0.1 degree.  At smaller radii it gives
the same contribution as the DM mass density profile by
\cite{Lokas2005} (see Fig.~\ref{fig:phicosmo_draco}). 
The Einasto profile, which predicts more mass at the intermediate radii which are resolved by the angular resolution of 0.1 degrees, gives the highest 
value of $\Phi_{\rm cosmo}$, while the NFW profile gives the same contribution
as \cite{Penarrubia2008}. The values of the results for the NFW and Einasto profiles depend on the mass adopted for the computation. We have used here
$10^9 \msun$ because the relative density profile was compatible with
the amplitude of the profile inferred by the data. \\

It is worth stressing here that the mass modeling of Draco produces only a
difference of a factor 2 to 3 in the flux predictions, while the indetermination arising from the unknown particle physics can sum up to several orders of magnitude.

\begin{figure}
   \includegraphics[width=0.5\textwidth]{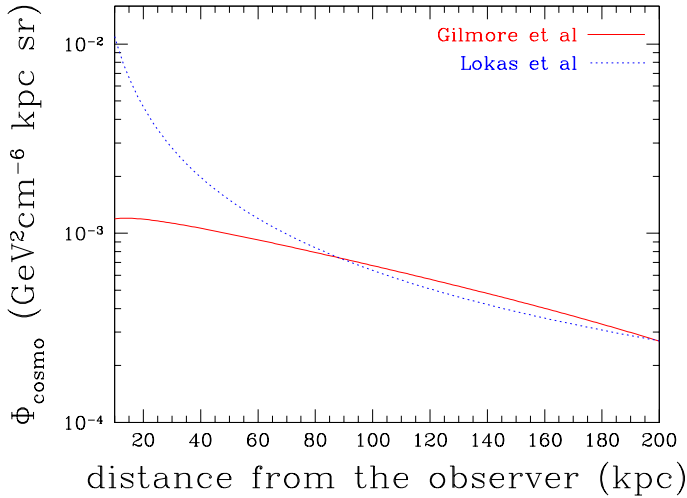}
\caption{The astrophysical/cosmological contribution $\Phi_{\rm
  cosmo}$ to the $\gamma$-ray flux computed in the centre of a
  Draco-like galaxy as a function of its distance to observer. The DM
  was considered to be radially distributed as in \cite{Gilmore2007} (solid line) and \cite{Lokas2005} (dotted)}
\label{fig:phicosmo_dracod}
\end{figure}

To investigate the reason why, e.g., the cuspy profile by \cite{Lokas2005}
and the cored profile by \cite{Gilmore2007} give the same value of
$\Phi_{\rm cosmo}$ towards the centre of Draco, we considered a
Draco-like dSph galaxy and changed its distance from the observer. We
then computed $\Phi_{\rm cosmo}$ toward the centre of the galaxy.
The result for the two profiles is plotted in 
Fig.~\ref{fig:phicosmo_dracod} as a function of
the imposed distance. The closer is the galaxy, the greater is the
contribution to $\Phi_{\rm cosmo}$ due to the cuspy radial profile of
the DM mass density. 
The geometrical acceptance of the Fermi-LAT detector is able to resolve 
the central cusp of the galaxy only if this is located at distances
smaller that 90 kpc. 
Further out, the two profiles give more or less the same result. 
Curiously enough,
the true location of the Draco dSph (80 kpc from us) lies exactly at
the border of this region, so that we can conclude that no matter
whether we choose either the cuspy DM profile by \cite{Lokas2005} or
the cored profile by \cite{Gilmore2007}, the estimate of the amount
of $\gamma$-rays expected from DM annihilation in the central region
of the galaxy will not change.

\begin{figure} 
   \includegraphics[width=0.5\textwidth]{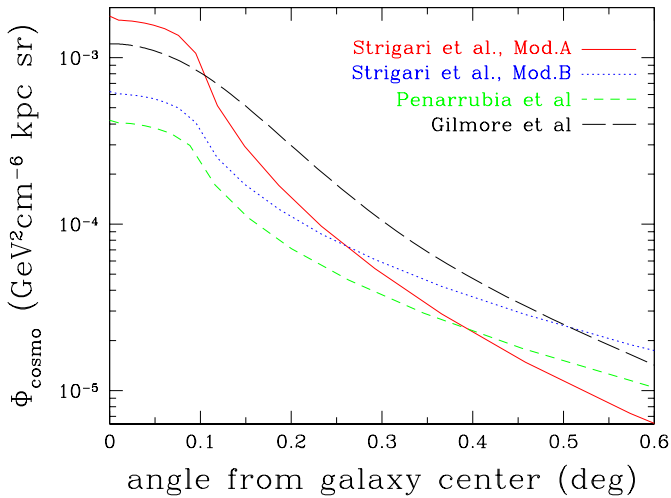}
\caption{The astrophysical/cosmological contribution $\Phi_{\rm
  cosmo}$ to the $\gamma$-ray flux derived as a function of the
  angular distance from the galaxy centre for Ursa Minor from the DM mass
  density radial profiles by \cite{Gilmore2007} (long-dashed line),
  \cite{Penarrubia2008} (dashed), and for the two fit to the NFW
profile proposed in \cite{Strigari2007} (solid and dotted)}
\label{fig:phicosmoUM}
\end{figure}

In Fig. \ref{fig:phicosmoUM} we plot the value of
 $\Phi_{\rm cosmo}$ for Ursa Minor for the cored \cite{Gilmore2007} profile,
as well as for the cuspy \cite{Penarrubia2008}, and for the two fit to the NFW
profile proposed in \cite{Strigari2007}. 
As in the case of Draco, the \cite{Penarrubia2008} profile gives the lowest 
value, while the two NFW models of \cite{Strigari2007} bracket the cored value at small angles. \\

\begin{figure}
   \includegraphics[width=0.5\textwidth]{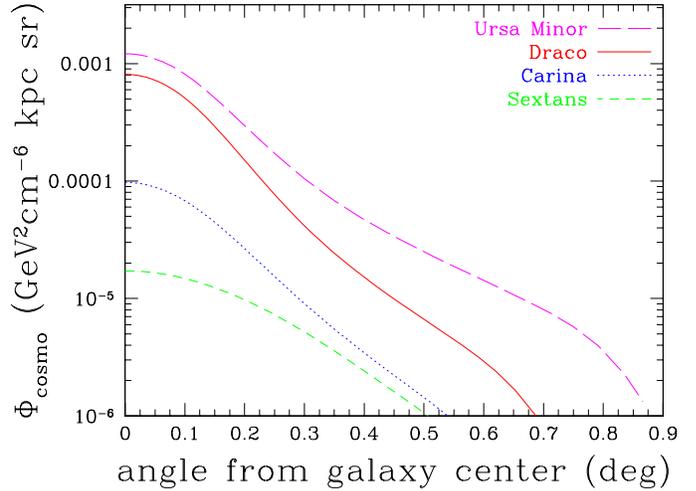}
\caption{The astrophysical/cosmological contribution $\Phi_{\rm
  cosmo}$ to the $\gamma$-ray flux derived as a function of the
  angular distance from the galaxy centre for Ursa Minor (long-dashed
  line), Draco (solid line), Carina (dotted line), and Sextans
  (short-dashed line) from the DM mass density
  radial profiles by \cite{Gilmore2007}.}
\label{fig:phicosmo}
\end{figure}

What has been afore-discussed holds for the other dSph galaxies
considered in this analysis too. 
As an example, in Fig.~\ref{fig:phicosmo} we plot the value of 
$\Phi_{\rm cosmo}$ obtained using the \cite{Gilmore2007} profile 
for the four dSph galaxies considered in this analysis.
The values obtained using cuspy profiles will not deviate significantly
from these values.

\begin{figure}
   \includegraphics[width=0.5\textwidth]{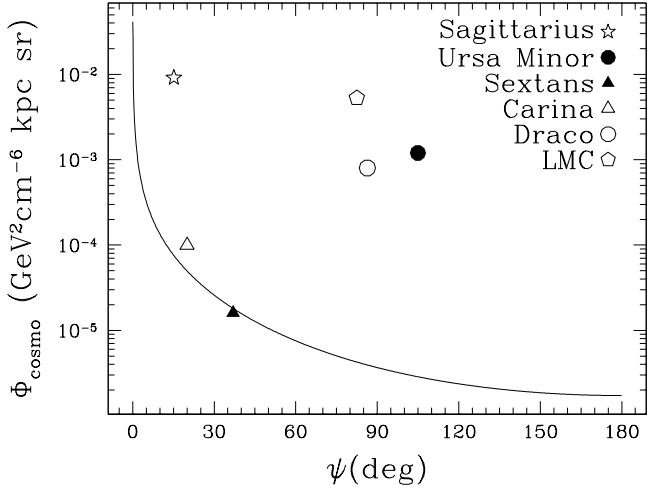}
\caption{The astrophysical/cosmological contribution $\Phi_{\rm
  cosmo}$ to the $\gamma$-ray flux derived as a function of the
  angular distance from the center of the Milky Way centre, computed for a
MW NFW smooth halo (solid curve), and for the central angular bin of  
Ursa Minor (filled circle), 
Draco (open circle), Carina (open triangle), and Sextans
  (filled triangle) derived using the DM mass density
  radial profiles by \cite{Gilmore2007}. Also superimposed are the values
for a NFW fit to the Sagittarius and the LMC galaxies.}
\label{fig:mw}
\end{figure}

Finally, Fig.\ref{fig:mw} shows the values of $\Phi_{\rm cosmo}$ 
obtained the profiles by \cite{Gilmore2007} and computed for
the LOS pointing toward the centre of the four dwarfs. These values
are compared to the curve obtained for the smooth
halo of the MW, obtained using Eq.\ref{eq:phicosmo}, 
an angular resolution of 0.1 degrees
and the NFW profile for the MW
($M_{\rm MW} = 10^{12} \msun$, $c=7.55$, $r_s=27.3 \kpc$).  
We observe that the dSph galaxies shine above the smooth Galactic 
halo at their position in the sky. Even more, we can say that, e.g. 
Draco is brighter than the Galactic halo at all angles larger than 2.3 
degrees above the Galactic center. \\
The central values of $\Phi_{\rm cosmo}$ for the Sagittarius dwarf galaxy 
and Large Magellanic Cloud (LMC) are shown in Fig.\ref{fig:mw} 
for a sake of comparison with those of the other dSph galaxies
we studied in detail.
The Sagittarius dwarf galaxy is located at a distance of about 24 kpc. 
Although it is heavily interacting with the Milky Way, it has a surviving 
stellar component thus it is likely to have a surviving dark matter halo. 
The observations suggest that it is dark matter dominated 
with a central stellar velocity dispersion of about 10 km s$^{-1}$ 
\cite{Ibata97}. 
According to recent observations and semi-analytic modelling 
(e.g. \cite{Strigari2008,Maccio2008}), the data consistent with all the 
DM halo of the dSph galaxies lie in the range between 20 and 40 km s$^{-1}$. 
We then modeled the inner regions of the 
DM halo of the Sagittarius dwarf with the same scale parameters as Draco 
(see \cite{Evans2004}) by assuming a NFW mass density profile and a mass 
of $M=10^9$ M$_\odot$. 
The LMC is located at about 50 kpc. 
We adopted for its DM halo the stripped NFW profile used 
by \cite{Tasi2004} ($M \sim 10^{10}$ M$_\odot$).

\section{Predictions for observation with the Fermi-LAT}
\label{sec:glast}

The map of the Fermi-LAT sensitivity to point sources of DM annihilations
has been obtained by \cite{Baltz2008} using the released
Fermi-LAT response functions.  
The sensitivity map was obtained for 55 days of observation and it shows
the minimum flux above 100 MeV which is necessary in order to achieve a
$5\sigma$ detection. The significance of the observed signal given the
local background counts is assigned with a maximum likelihood analysis
assuming Poisson statistics.
\cite{Baltz2008} found that the sensitivity
suffers very little dependence from the underlying particle physics.
Therefore, the obtained values can be considered valid as long as the
source appears point like in the sky, that is as long as its angular
size does not exceed 0.25 degrees.
As it can be seen in Fig.~\ref{fig:phicosmo}, the $\gamma$-ray flux
expected from our dSph galaxies decreases by almost one order of
magnitude at the angular distance of 0.25 degrees from the galaxy
centre. For this reason, we can assume they are
point-like sources and use the results of \cite{Baltz2008}
for reference.
Draco, Ursa Minor, and Sextans lie in a region of the sky where
the $5\sigma$ detection flux above
100 MeV is $1.5 \times 10^{-8}$ \phcms in $\sim$ 2 months. This translates into
$\phi^{\rm 1 yr}_{5 \sigma} = 6 \times 10^{-9}$ \phcms\ in 1 year
of data taking and in the units we used throughout this paper. In the case of
Carina, it is $\phi^{\rm 1 yr}_{5 \sigma}=8 \times 10^{-9}$ \phcms ,
since the galaxy is closer to the Galactic plane.

If we consider the best value for $\Phi_{\rm PP}$ ($>$ 100 MeV) from
Fig.~\ref{fig:phisusy} ($\Phi_{\rm PP} \sim 6 \times
10^{-8}\;\cm^4\;\kpc^{-1}\;\GeV^{-2}\;\sec^{-1}\;\sr^{-1}$)
and the average value of $\Phi_{\rm cosmo}$ toward the
galaxy centre ($\psi = 0$) from Fig.~\ref{fig:phicosmo_draco} and
Fig.~\ref{fig:phicosmoUM}, we end up
with the following best-particle-physics-case estimates 
for the $\gamma$-ray flux from
DM annihilation in Draco: 
\begin{equation}
\Phi^{\rm Draco}_{\gamma} (> 100 \MeV) = (4.6 \pm 1.1) \times 10^{-11} \  
 {\rm ph} \cm^{-2} \sec^{-1}
\end{equation}
and Ursa Minor:
\begin{equation}
\Phi^{\rm UMin}_{\gamma} (> 100 \MeV) = (6.0 \pm 3.8) \times 10^{-11} \ 
 {\rm ph} \cm^{-2} \sec^{-1} \, .
\end{equation}
The error is given by the standard deviation for the values of
$\Phi_{\rm cosmo} (\psi=0)$ obtained using different DM density profiles inferred
by dynamical modeling.
We therefore do not focus on a specific profile when giving the
value of the expected $\gamma$-ray flux. Indeed, our result
is obtained by averaging over different fits to the data. It is worth noticing
that predictions made using profiles inferred by astronomical data are
robust within a 60 \% relative error which is expected while changing fit. 
This
means that they can provide a reliable order-of-magnitude estimate of the
real flux.
We will not further consider the case of Carina and Sextans since they
give a lower flux.
Yet, the calculations of the expected $\gamma$-ray fluxes from
DM annihilation in these galaxies are straightforward.

Even in the case of Draco and Ursa Minor the upper value of the predicted flux 
within the error ends up to be 2 orders of magnitude below  
that required for
detection in 1 year of data taking with the Fermi-LAT, referring to  
\cite{Baltz2008}. 
This means that there is no hope of detection unless we allow for the presence of boost factors. \\
Though brigthest than the dSph considered in this analysis, 
none of the Sagittarius dwarf galaxy and LMC have a predicted flux 
which could be detected with the Fermi-LAT.

\begin{figure}
   \includegraphics[width=0.5\textwidth]{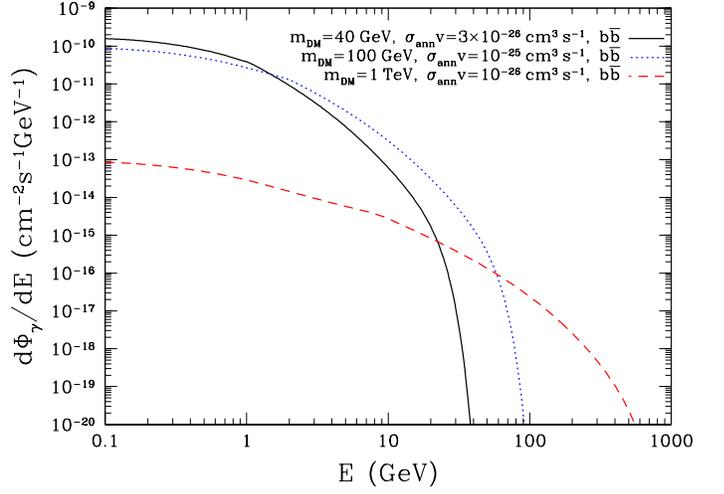}
\caption{Differential $\gamma$-ray fluxes as a function of the energy, expected from DM annihilation in the center of Ursa Minor. Fluxes are computed using the best value for $\Phi_{\rm cosmo}$ given by model A of \cite{Strigari2007},
 and are presented for a 40 GeV (solid line), 100 GeV (dotted) and 1 TeV (dashed) DM particle annihilating into $b \bar b$.}
\label{fig:flussi}
\end{figure}

In Fig. \ref{fig:flussi} we show the differential $\gamma$-ray fluxes  
expected from DM annihilation in the center of Ursa Minor for
a 40 GeV, 100 GeV and 1 TeV DM particle annihilating into $b \bar b$. 
Fluxes are computed using the best 
value for $\Phi_{\rm cosmo}$ given by model A of \cite{Strigari2007}.
The values of $\sigma_{\rm ann} v$ have been chosen 
as in Fig.\ref{fig:phisusy}. \\

We note that if we use the same values for the
annihilation cross-section and for the mass ($\sigma_{\rm ann} v = 5
\times 10^{-26}$ \cms, $m_\chi$ = 46 GeV) as in
\cite{Strigari2007}, as well as their model A for the density profile,
we find a prediction for Ursa Minor which is $\sim10$ times smaller
than their value. In fact, we get $\Phi_{\gamma} (> 5 \GeV) \sim 2.5 \ [5.4]
\times 10^{-12} \ {\rm ph} \cm^2 \sec^{-1}$ for annihilation into $b \bar b$ 
[$\tau^+ \tau^-$] 
to be compared with their
value $\Phi_{\gamma} (> 5 \GeV) \sim 3 \times 10^{-11} \ {\rm ph}
\cm^2 \sec^{-1}$. This is due to the over-estimated number of photon 
yields above 5 GeV ($\int_{5 \ GeV}^{m_\chi} dE dN_\gamma / dE_\gamma \sim$ 4.2) which is derived in their paper. We found a number of photon yields which is an order of magnitude smaller both using the \cite{Fornengo2004} and the \cite{Bergstrom1998} parametrization for $dN_\gamma / dE_\gamma$, the latter being the one used by \cite{Strigari2007}. 

Investigating possible sources of astrophysical boost factors becomes 
necessary in order to understand the feasibility of a DM signal detection with 
the Fermi-LAT. To this purpose, in the following sections we account for the 
effect of the presence of clumps or of a SBH inside the dSph
galaxies we are considering.

\subsection{Boost factor due to the presence of dark matter clumps}
\label{sec:bfclumps}

According to the CDM scenario, each halo formed through the merging
and accretion of smaller haloes, which still survive and orbit inside
the larger one. The minimum mass of these subhaloes is $\sim 10^{-6}$
M$_\odot$ according to analytical estimates
(\cite{Green2004,Green2005}).
High-resolution $N-$body experiments, though they stop at high redshift
($z=26$), are able to resolve field haloes as small as $\sim 10^{-6}$
M$_\odot$.  Their mass function is well approximated by a power law
\begin{equation}  
{\rm d} n(M)/{\rm d ln}(M) \propto M^{-\alpha},
\label{eq:mass}  
\end{equation}  
with $\alpha$ = 1, independently of the mass of the host halo, over
the large redshift range between 0 and 75 and the mass interval
between about $10^{-6}$ and $10^{10}$ M$_\odot$
(\cite{Diemand2005,Giocoli08}).

Assuming that the radial distribution of subhaloes traces that of the
host galaxy, we can model the number density of subhaloes per unit
mass at a distance $R$ from the galaxy centre as
\begin{equation}  
\rho_{\rm sh}(M,R) = A M^{-2} \theta (R - r_{\rm min}(M)) \rho_{\rm gal}(R) 
\ {\rm M}_\odot^{-1}  \kpc^{-3} , 
\label{eq:rho}  
\end{equation}  
where $A$ is a normalization factor which takes into account the
hypothesis that $10\%$ of the Milky Way (MW) mass is distributed in
substructures with masses in the mass range between $10^7$ and
$10^{10}$ M$_\odot$
(\cite{Diemand2005}.)
The effect of tidal disruption is accounted for by the Heaviside step
function $\theta (r - r_{\rm min}(M))$, where the tidal radius $r_{\rm
min}(M)$ is computed according to the Roche criterion as the minimum
distance at which the subhalo self-gravity at its scale radius equals
the gravity pull of the halo host computed at the orbital radius of
the subhalo.
The debate on the survival and partial disruption of these haloes is
still open, and many issues are still unsolved, such as the true mass
function after tidal interactions in the host halo, the inner
structure and concentration of the subhaloes themselves. We refer to
\cite{Pieri2007} for a detailed discussion of the problem.

Once we assumed a model for the subhalo population, the boost factor due
to the presence of clumps distributed according to $\rho_{\rm
sh}(M,r)$ is computed as the ratio of the integral over the galaxy
volume of the density squared including subhaloes to the same integral
for the smooth galaxy only:
\begin{equation}
{\rm BF}_{\rm SH} = \frac{\int_{\rm gal} dV  \rho_{\rm gal,sm}^2 + 
  \int_{\rm gal} \int_{\rm M_{sh}}dV dM \rho_{\rm sh} \int_{\rm halo} dV  
  \rho_{\rm h}^2 }{\int_{\rm gal} dV \rho_{\rm gal,sm}^2}
\label{eq:bfclumps}
\end{equation}
where $\rho_{\rm gal,sm}$ is the smooth profile of the DM component of
the host halo which is not virialized into clumps and $\rho_{\rm h}$
is the internal DM density profile of each subhalo.

\cite{Pieri2007} found a relationship between the different subhalo
models leading to more or less impressive boost factors for the MW,
the total number of photons produced at high galactic latitudes by the
annihilation of DM particles in all the subhaloes falling into a given
cone of view (of the order of $10^9$), the EGRET measurement of the
extragalactic $\gamma$-ray background (EGB), and the allowed particle
physics contribution.
They observed how a given model for the subhalo population can not
predict a number of photons greater than those observed by EGRET at
high latitudes, where the $\gamma$-ray flux is thought to have a
diffuse origin. Consequently, a maximum number of predictable photons
exists. This means that the two factors $\Phi_{\rm PP}$ and $\Phi_{\rm
cosmo}$ must be tuned in order not to exceed the EGRET limit.  In the
most optimistic case, they will be tuned as to give exactly the number
of photons observed by EGRET. This means that, if we assume a subhalo
model for the MW, the value of $\Phi_{\rm PP}$ can be shifted down or
up to match the EGRET level 
(up to the level of the best-particle-physics case of Fig.\ref{fig:phisusy}).
Now, in the lack of accurate
models which account for the presence of subhaloes inside subhaloes, 
we make the simplifying assumption that the subhalo
population of a dSph galaxy is described by the same
subhalo model which we have assumed to be valid in the MW, so that the
restrictions on $\Phi_{\rm PP}$ must still hold.

\begin{figure} 
   \includegraphics[width=0.5\textwidth]{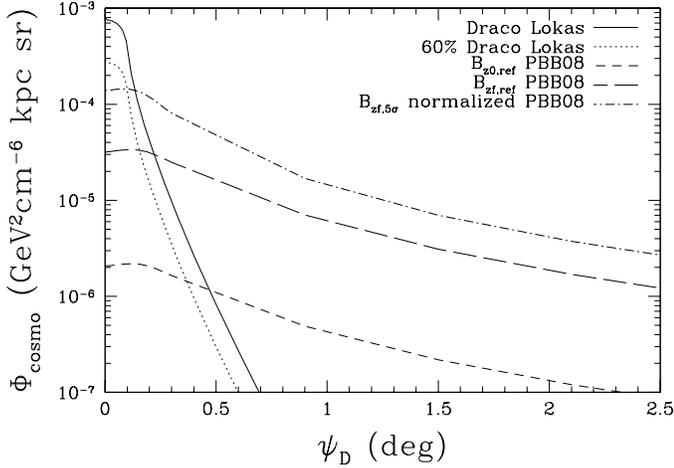}
\caption{The astrophysical/cosmological contribution $\Phi_{\rm
  cosmo}$ to the $\gamma$-ray flux as a function of the
  angular distance from the galaxy centre, derived for Draco using the DM mass
  density radial profiles by \cite{Lokas2005} (solid line).
  Dashed, dot-dashed and long-dashed lines show the contribution of all 
  sub-subhaloes.
  The dotted line shows the contribution of the DM halo of Draco when 
  only 60\% of its mass is smoothly distributed in the halo.}
\label{fig:bfclumpsdraco}
\end{figure}

We have computed the boost factors in the cases of Draco and Ursa Minor,
for all the subhalo models considered in \cite{Pieri2007}, 
(PBB08 in Fig.\ref{fig:bfclumpsdraco}) using Eq. \ref{eq:bfclumps}. 
We found that the values for the boost factors range from 1.6 to 850, but when
applying the corresponding limits on $\Phi_{\rm PP}$, we end up, for
any clump model, with an estimate of the maximum flux which may be produced by
the clumps in Draco and Ursa Minor which is still compatible with the
EGRET EGB and with the constraints given by particle physics shown in Fig.
\ref{fig:phisusy}. 
The overall maximum enhancement of the flux obtained using 
Eq. \ref{eq:bfclumps}, once scaled for the EGRET limit, is of a factor 70. \\

The boost factor due to the presence of subhaloes has been 
computed analytically
also in \cite{Strigari2007} and \cite{Kuhlen2008}. 
The overall values
is of the same order of magnitude of the one we obtain using the 
$B_{\rm z0,ref}$
model of \cite{Pieri2007}. For that specific model, we obtain 
${\rm BF}_{\rm SH}$ = 2, which is not expected to give an enhancement
of the $\gamma$-ray flux which could be significant for detection. \\

The values of the boost factors obtained using Eq. \ref{eq:bfclumps}, as 
well as those obtained analytically in \cite{Strigari2007} 
and \cite{Kuhlen2008},
hold when integrating over all the galaxy. They are thus the numbers to 
consider when the galaxy is so far as to be pointlike inside the detector 
acceptance. 
This is indeed not the case for the nearby 
dwarfs considered in this analysis.\\

To understand what could really be the effect of sub-subhaloes in the closest
dwarfs, we assumed a NFW profile for the substructures, defined 
by the concentration parameter $c$ distributed according to a log-normal 
probability $P(c)$, and have computed their contribution to the annihilation 
signal as in \cite{Pieri2007}:
$$ 
\Phi^{\rm cosmo}(\psi, \Delta \Omega) = \int_M d M \int_c d c \int \int_{\Delta \Omega}
d \theta d \phi \int_{\rm l.o.s}  d\lambda
$$
$$
[ \rho_{sh}(M,R(\rsun, \lambda,\psi, \theta, \phi)) \times P(c) \times
$$
\begin{equation}
\times \Phi^{\rm cosmo}_{\rm halo}(M,c,r(\lambda, \lambda ', \psi,\theta ', \phi ')) \times J(x,y,z|\lambda,\theta, \phi) ]
\label{smoothphicosmo}
\end{equation}
where $\Delta \Omega$ is the solid angle
of observation pointing in the direction of observation $\psi$ and defined 
by the angular resolution of the detector $\theta$; $\rho_{\rm sh}$ is the 
sub-subhaloes mass and distribution function inside the dwarf; 
$J$ is the Jacobian determinant; $R$ is the galactocentric distance, 
$r$ is the radial coordinate inside the single sub-subhalo located 
at distance $\lambda$ from the observer along the line of sight defined 
by $\psi$ and contributing to the diffuse emission;
$\Phi^{\rm cosmo}_{\rm halo}$ describes the emission from each sub-subhalo. 
As in the case of the MW,  $\rho_{\rm sh}$ is normalized such that 
$10\%$ of the Draco mass is distributed in
substructures with masses in the mass range between $10^{-5}$ and
$10^{-2}$ M$_{\rm Draco}$. When integrating over all sub-subhaloes, we end up 
with 40 \% of the Draco mass distributed in $10^{12}$ sub-subhaloes. 
The results of the computation of $\Phi^{\rm cosmo}$ 
using Eq.\ref{smoothphicosmo} are shown in Fig.\ref{fig:bfclumpsdraco} 
using different models for the concentrations parameters, namely the 
$B_{\rm z0,ref}$, $B_{\rm zf,ref}$ and $B_{\rm zf,5 \sigma}$  described in 
\cite{Pieri2007}.
Superimposed are the values of $\Phi^{\rm cosmo}$ for the profile 
by \cite{Lokas2005} with 100\% and 60\% of the mass of Draco smoothly 
distributed in the halo. The sum of the latter contribution and the 
sub-subhalo ones should be compared with the 100\% smooth halo (solid line).
We note that the effect of adding sub-substructures can give an 
enhancement of several orders of magnitude at large angles, 
where the overall flux is though too low to be detected, 
even in the presence of sub-subhaloes. 
Yet, the effect in the very inner parts of the halo will be that of decreasing 
the expected signal,
and corresponding boost factor defined as $(\Phi^\gamma_{\rm 60\% smooth} + 
\Phi^\gamma_{\rm sub-subhaloes}) \/ \Phi^\gamma_{\rm 100\% smooth}$ is less than 1 
where the larger detectable flux is expected, that is toward the galaxy center.

\subsection{Boost factor due to the presence of a black hole}
\label{sec:bfbh}

So far there are only two examples of dwarf galaxies suggested to host
a SBH. \cite{Maccarone2005} discussed the possibility that the radio
source near the core of the Ursa Minor dwarf galaxy is a SBH. They
give a mass upper limit of $\sim10^4$ M$_\odot$. 
\cite{Debattista2006} assumed that the double nucleus of the dE
VCC~128 is a disk orbiting a SBH. They derived a SBH mass of
$\sim10^7$ M$_\odot$.
The lack of SBHs in dwarf galaxies was explained by
\cite{Ferrarese2006}. They found that for galaxies less massive than
few $10^9$ M$_\odot$ the formation of a compact stellar nucleus is
more likely than that of a SBH. Both stellar nuclei and SBHs contain a
mean fraction of about $0.2\%$ of the total mass of the galaxy. The
same conclusion was reached by \cite{Wehner2006} and
\cite{Cote2006}.

Nevertheless, a value for the SBH mass of the dSph galaxies studied in
this paper can be inferred using the $M_{\rm SBH}-\sigma$ relation
(see \cite{Ferrarese2005} for a review). Extrapolating the 
scaling law defined by SBHs in massive galaxies to the constant
$\sigma\approx10$ \kms\ measured in the sample galaxies, the derived
SBH mass is $M_{\rm SBH} \approx10^2$ M$_\odot$.

\cite{Gondolo1999} and \cite{Merritt2004} studied the effect of an
adiabatically accreted SBH on a cuspy DM mass density profile
[$\rho(r) \propto r^{-\gamma}$ with $0 < \gamma < 2$]. They found that
the SBH induces a central density spike described by a power-law
radial profile with index
\begin{equation}
\gamma_{\rm s}=\frac{9-2\gamma}{4-\gamma}
\label{eq:gammaspike}
\end{equation}
over a region of radius 
\begin{equation}
r_{\rm s} \simeq 0.2 \ r_{\rm SBH} = 0.2 \frac{G M_{\rm SBH}}{\sigma^2}
\label{eq:rspike}
\end{equation}
where $r_{\rm SBH}$ is the radius of gravitational influence of a SBH
with a mass $M_{\rm SBH}$ and $\sigma$ is the DM velocity dispersion
at $r_{\rm SBH}$. Eq.~\ref{eq:gammaspike} reflects the condition of
adiabaticity, requiring the SBH formation time to be much larger than
the dynamical timescale at a distance $r_{\rm SBH}$.

Assuming the cuspy DM mass density profile ($\gamma=1$) given by
\cite{Lokas2005} for Draco, it results
\begin{equation}
\rho_{\rm SBH+DM}(r) = \left\{
  \begin{array}{ll}
    \rho ( r_{\rm c} ) & 
        \mbox{ $r \leq  r_{\rm c}$} \\
    \rho ( r_{\rm s}) \left( \frac{r}{r_{\rm s}} \right)^{-7/3} & 
        \mbox{$r_{\rm c} < r \leq   r_{\rm s}$}\\
    \frac{M_{\rm DM}}{4 \pi r^2_{\rm b} r} \exp{\left( -\frac{r}{r_{\rm b}} \right)} &
        \mbox{$r > r_{\rm s}$}
\end{array}
\right.
\label{rhobh}
\end{equation}
where $M_{\rm DM}$ and $r_{\rm b}$ are the total DM mass and break
radius as in \cite{Lokas2005}, respectively. A core radius $r_{\rm c}
= 10^{-8}$ kpc was imposed to prevent the high annihilation rate which
would destroy the spike. 
On the other hand, a mass density profile with constant-density core
($\gamma=0$) would not allow the growth of the spike. Therefore, it
was not considered.

\begin{figure}
   \includegraphics[width=0.5\textwidth]{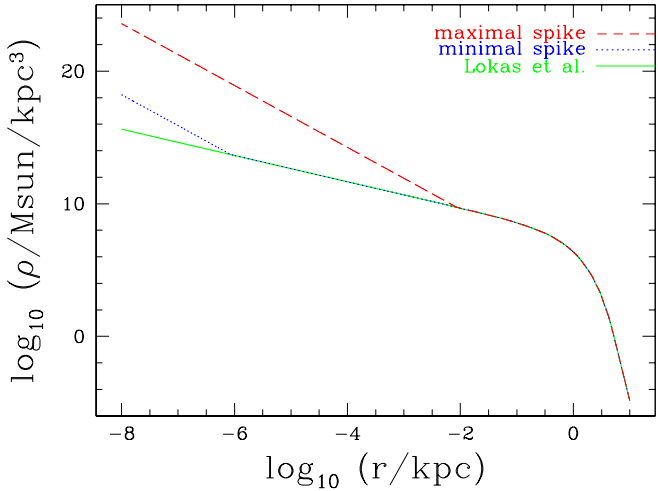}
\caption{The density spike induce by a SBH on the DM mass 
  density profile of Draco. The initial profile is taken from
  \cite{Lokas2005} (solid line). The dotted and dashed lines
  correspond to a final density profile with a spike radius $r_{\rm
  s} = 10^{-3}$ and $10$ pc, respectively.} 
\label{fig:densitydracospike}
\end{figure}

The effect of a SBH on the DM mass density profile by
\cite{Lokas2005} is shown in Fig.~\ref{fig:densitydracospike}
for two extreme values of the $r_{\rm s}$. 
For $\sigma = 10$ \kms , $r_{\rm s} = 10^{-3}$ pc corresponds to
$M_{\rm SBH} \simeq 10^2$ M$_\odot$ and $r_{\rm s} = 10 $ pc
corresponds to $M_{\rm SBH} \simeq 10^6$ M$_\odot$. The adopted value
of $\sigma$ is consistent with the stellar velocity dispersion
measured in the dSph galaxies we studied (\cite{Gilmore2007}). It
represents an upper limit for the DM velocity dispersion, since the
stars are a tracer population of both the luminous and DM components.

The boost factor due to the presence of a SBH is 
\begin{equation}
{\rm BF}_{\rm SBH} = \frac{\int_{\rm gal} dV  
  \rho_{\rm SBH+DM}^2}{\int_{\rm gal} dV \rho_{\rm gal,sm}^2} .
\label{eq:bfbh}
\end{equation}
Of course, a different parametrization of the SBH properties would result in
different values of ${\rm BF}_{\rm BH}$. In particular, if we assume $r_{\rm
s} = 10^{-3}-10^{-2}$ pc we end up with a boost factor in the range
$[1,10]$, while if we consider $r_{\rm s} = 1-10$ pc the boost factor
becomes of the order of $[5 \times 10^7,10^9]$.

A SBH mass $M_{\rm SBH} \approx10^2$ M$_\odot$ results in $r_{\rm s} =
10^{-3}$ pc, and in a boost factor equal to 1. On the other hand,
\cite{Gilmore2007} the existence of observed small stellar clumps
inside Ursa Minor is not compatible with the presence of a spike by
which they would have been tidally destroyed.

The MAGIC collaboration has recently published (\cite{MAGIC}) 
an upper limit on the
$\gamma$-ray flux above 140 GeV for Draco at $10^{-11}$ \phcms.  
If we multiply the value of $\Phi_{\rm cosmo}$
in the direction of the centre of Draco by a boost factor $\rm BF_{\rm
BH} = 5 \times 10^7$ we end up with $\Phi_{\rm cosmo}|_{\rm BF_{\rm
BH}} = 4 \times 10^4\;\GeV^2\;\cm^{-6}\;\kpc\;\sr$. This means that,
if such a boost factor exists, either the DM particle mass is below
140 GeV and its annihilation products could not be observed with
MAGIC, or that $\Phi_{\rm PP} < 2.5 \times
10^{-16}\;\cm^4\;\kpc^{-1}\;\GeV^{-2}\;\sec^{-1}\;\sr^{-1}$ which is a
very low (though allowed) value.

\section{Conclusions}
\label{sec:conclusions}

The Fermi-LAT telescope was launched in June 2008 and is taking 
data on $\gamma$-rays in
the energy range between about 20 MeV and 300 GeV. Its all-sky survey
operation mode will allow an unprecedented precise study of the
$\gamma$-ray sky, so that many DM models will be tested. 
The dSph galaxies of the Local Group will be primary targets for DM
analysis with the Fermi-LAT, because of the low astrophysical background 
expected in their direction.
Therefore, we studied the detection limit of the $\gamma$-ray flux
from DM annihilation in four of the nearest dSph galaxies, namely
Draco, Ursa Minor, Carina, and Sextans.

State-of-art DM density profiles were available for these galaxies and
we computed the expected $\gamma$-ray flux from DM annihilation for
different particle physics parameters. We varied the DM particle mass
as well as the annihilation cross-section and branching
ratios. 
%
We found that the presence of NFW-like cusp or constant density core
in the DM mass density profile does not produce any relevant effect in
the $\gamma$-ray flux due to a combination of the geometrical
acceptance of the Fermi-LAT detector, which is not able to resolve the
very inner shells of the studied galaxies, and the distance of
the sample galaxies.
 
In the case of Draco and Ursa Minor, we found that they would shine above
the Galactic smooth halo for all but the smallest angles ($\sim$ 2 degrees) 
above the Galactic Center. Yet,
the upper values of the predicted flux 
were found to be about two orders of magnitude below the Fermi-LAT 
detection threshold as derived in \cite{Baltz2008}.
Such values were
computed for the most optimistic particle physics scenario of a 
40 GeV particle with  $\sigma_{\rm ann} v= 3 \times 10^{-26}$ \cms
annihilating into $b \bar b$.
We have shown how the effect of the boost factor due the presence of a 
population of DM
clumps inside the dSph galaxies, though possibly dramatic (of the
order of $10^3$ when integrated over the whole galaxy),
 had to be rescaled for the limit on the EGB measured by
EGRET. The overall maximal effect was reduced to a factor 70.
The reader should keep in mind though that the calculation was made for 
a toy-model where the subhalo population of the
dwarf galaxies is described by the same model used for the MW.
In any case, since the closest dwarfs are not pointlike for the 
Fermi-LAT angular 
acceptance $\Delta \Omega$, the factor to be taken into consideration is the 
effect of the sub-subhaloes inside $\Delta \Omega$, which resulted in a 
decrease 
of the expected flux due to a redistribution of the DM inside the halo.
The presence of a central SBH in agreement with the $M_{\rm SBH}-\sigma$
relation extrapolated to the observed low $\sigma$ values resulted in
a negligible boost factor. More extreme models would result in a much
higher boost factor, though they are not theoretically supported.

Contrarily to previous papers which addressed the presence of subhaloes 
or of SMBHs to boost the signal, we have demonstrated that 
the boost factor must be searched for in some exotic extension or 
modification of the particle physics sector.
Unless such a scenario happens, 
we conclude that the annihilation of DM inside the local dwarfs is unlikely
to be detected with the Fermi-LAT.

As a further development, it will be interesting to repeat the
present analysis of boost factors for the recently catalogued
ultra faint dwarfs of the Local Group. \cite{Strigari2008} 
have computed the expected $\gamma$-ray flux from those sources, deriving
the halo parameters from kinematical data. The inclusion of such
galaxies in our study will improve the sensitivity of 
a joint multi-centred likelihood analysis. 

It is worth noticing that,since the DM spectra would be the
same for all the galaxies, such an analysis 
could be performed in order to maximise the detection efficiency and
to allow portions of the particle physics phase-space to be
explored.

\section*{Acknowledgments}
\noindent 
We acknowledge Torsten Bringmann for valuable discussion and suggestions.
We would like to thank also Giovanni Busetto, 
Sergio Colafrancesco, Andrea Lionetto, Mos\`e Mariotti
and Riccardo Rando for discussion and comments, and Mark
Wilkinson for sending us his data about Carina and Sextans.  This work
was made possible through grants PRIN 2005/32 by Istituto Nazionale di
Astrofisica (INAF) and CPDA068415/06 by Padua University
and was carried out under contract ASI/INAF I/010/06/0.

\end{document}